\documentclass{aa}
\usepackage{psfig}

\def\amin{\ifmmode ^{\prime}\else$^{\prime}$\fi}
\def\farcm{\hbox{$.\mkern-4mu^\prime$}}    
\newbox\grsign \setbox\grsign=\hbox{$>$}
\newdimen\grdimen \grdimen=\ht\grsign
\newbox\laxbox \newbox\gaxbox
\setbox\gaxbox=\hbox{\raise.5ex\hbox{$>$}\llap
     {\lower.5ex\hbox{$\sim$}}}\ht1=\grdimen\dp1=0pt
\setbox\laxbox=\hbox{\raise.5ex\hbox{$<$}\llap
     {\lower.5ex\hbox{$\sim$}}}\ht2=\grdimen\dp2=0pt

\def\lax{$\mathrel{\copy\laxbox}$}

\def\am{AM~Her}
\def\ex{EX~Hya}
\def\axj{AX\,J2315$-$592}
\def\asc{{\it ASCA}}
\def\gin{{\it Ginga}}
\def\mnras{{MNRAS}}
\def\apj{ApJ}
\def\apjl{ApJ}
\def\apjs{ApJ Suppl}
\def\aaa{A\&A}

\def\pasj{PASJ}

\def\ka{$K\alpha$}

\def\etal{{\it et al.}}
\def\iue{{\it IUE}}
\def\ein{{\it Einstein}}
\def\exo{{\it EXOSAT}}
\def\ro{{\it ROSAT}}
\def\rxj{RX\,J1802.1+1804}

\begin{document}

   \thesaurus{06        
              (02.01.2, 
               08.02.3, 
               08.09.2, 
               08.14.2, 
               13.25.5  
                         )}

\title{ASCA Observation of the polar RX\,J1802.1+1804}

\author{M. Ishida \inst{1} 
       \and J. Greiner \inst{2}\thanks{Present address: Astrophysical Institute
        Potsdam, An der Sternwarte 16, 14482 Potsdam, Germany; jgreiner@aip.de}
       \and R.A. Remillard \inst{3}
       \and C. Motch \inst{4} }

\offprints{M. Ishida}

   \institute{Institute of Space and Astronautical Science,
              3-1-1 Yoshinodai, Sagamihara, Kanagawa 229, Japan \\
          email: ishida@astro.isas.ac.jp
          \and Max-Planck-Institut f\"ur extraterrestrische Physik,
               85740 Garching, Germany 
          \and Center for Space Research, MIT, 77 Mass Ave, Cambridge, 
                MA 02139 U.S.A. \\
                email: rr@space.mit.edu
          \and Observatoire de Strasbourg, 11 rue de
                l'Universit\'e, F-67000 Strasbourg, France \\
                email: motch@astro.u-strasbg.fr
          }

   \date{Received 22 December 1997; accepted 14 May 1998}

\maketitle

\begin{abstract}

We present X-ray data of {\rxj} obtained by {\asc}.
Although it shows a clear orbital intensity modulation with
an amplitude of nearly 100\% below 0.5~keV in {\ro} data,
the {\asc} light curves are nearly flat except for a possible dip
lasting about one-tenth of the orbital period.
We discuss this within the model assumption of a stream-eclipsing geometry
as derived from the ROSAT observations.

The {\asc} X-ray spectrum can be represented by a two temperature
optically thin thermal plasma emission model with temperatures of
$\sim$1\,keV and $>$7\,keV, suggesting postshock cooling as observed in {\ex}.
A remarkable feature of the spectrum is the strong iron {\ka} emission line
whose equivalent width is $\sim$4\,keV.
To account for this, an iron abundance of greater than 
at least 1.3 times Solar is required. 
A combined spectral analysis of the {\ro} PSPC and {\asc} data indicates that
the $N_{\rm H}$-corrected
flux ratio of the soft blackbody (0.1--2.4~keV) to the hard
optically thin thermal plasma emission (2--10~keV) is as large as $\sim 10^4$.


\end{abstract}

\section{Introduction}

A polar (or {\am} type object; Cropper 1990) 
is an accreting binary composed of a mass-donating
low mass secondary star and a magnetized white dwarf with a field strength of
the order of 10-100\,MG.
Matter from the secondary accretes along the field lines onto a small region
of the white dwarf close to the magnetic pole.
Since the flow is highly supersonic, a standing shock is formed close to
the white dwarf, and a hot plasma with a temperature of $10^8$~K is 
formed. From the postshock plasma, optical cyclotron emission and optically 
thin thermal plasma emission in X-rays have been observed.
In addition to this, blackbody emission with a temperature of 10--40~eV
is observed.
Although the blackbody component is considered to be radiated from the surface
of the white dwarf around the postshock plasma via reprocessing of the 
cyclotron and the hard X-ray radiation,
its intensity is usually much larger than that of the cyclotron and
hard X-ray radiation. This has become known as the so-called 
`soft excess problem' (Rothschild \etal\ 1981).
Beuermann and Burwitz (1995) recently suggested 
that the amount of the soft excess is correlated with
the strength of the magnetic field of the white dwarf.

It is known that the hard X-ray continuum spectrum of magnetic cataclysmic 
variables (mCVs) can, to a first approximation, be fitted by
an optically thin thermal plasma emission spectrum with a single temperature
undergoing photoelectric absorption represented by a single hydrogen column 
density. To represent the {\exo} spectra of intermediate polars,
Norton and Watson (1989) introduced the so-called
`partial-covering absorber model'
in which photoelectric absorption was represented by two column densities.
Complex absorption was found also in polars with {\gin} observations
(Ishida and Fujimoto 1995).

It has also been expected that the hard X-ray emitting hot plasma
is gradually cooled by cyclotron emission and bremsstrahlung
(Aizu 1973, Frank, King and Lasota 1983, Imamura and Durisen 1983),
and that the hard X-ray spectrum consists of multi-tem\-pe\-ra\-ture emission
components.
Such a multi-temperature emission spectrum was first suggested by an {\ein}
observation of {\ex} (Singh \& Swank 1993), and later was established by
an {\asc} observation
(Ishida, Mukai and Osborne 1994, Fujimoto and Ishida 1997).
Note, however, that it is only for {\ex} that the existence
of multi-temperature emission is confirmed observationally.

{\rxj} was discovered during the search for supersoft X-ray sources in the
{\ro} all-sky survey data (Greiner, Remillard and Motch 1995).
Greiner, Remillard and Motch (1995, 1998) have analyzed all the {\ro}
data taken between 1990 September and 1993 Sep\-tem\-ber, 
and found a coherent period of
0.07847977(11)~d ( = 1.8835145$\pm$0.000003~hr).
The pulse profile in the band $<0.5$~keV is characterized by a deep intensity
minimum, with basically no X-ray flux, lasting 0.1 orbital phase.
The X-ray spectrum is characterized by  strong blackbody emission
with a temperature of $20\pm 15$~eV, 
with a clear excess emission above 1~keV, which has been approximated by
thermal bremsstrahlung with a temperature of 20~keV.
The absorption-corrected 0.1--2.4~keV fluxes of
the blackbody and the thermal brems\-strahlung components are 
$7 \times 10^{-11}$~erg~cm$^{-2}$s$^{-1}$ and 
$8 \times 10^{-13}$~erg~cm$^{-2}$s$^{-1}$, respectively,
suggesting a huge soft excess of nearly 90 in the 0.1--2.4 keV band.
Szkody {\etal} (1995) carried out photometry, spectroscopy and polarimetry
in the optical band, and found several characteristics of polars
such as He{\small II} emission line stronger than H$\beta$ and
circular polarization of 4\%.
All these properties strongly indicate that {\rxj} is a polar.

In this paper, {\asc} data of {\rxj} taken in 1996 Sep-Oct are presented.
In \S~2 we describe how the {\asc} observation was carried out.
In \S~3 light curve and spectral analysis are presented.
We discuss these properties in \S~4 in combination with
the {\ro} and {\iue} data. In \S~5 we summarize our results.

\section{Observation}

The {\asc} observation of {\rxj} was carried out between 1996 September
30.75 and October 2.96 (UT). {\asc} is equipped with four equivalent 
X-ray Telescopes (XRT: Serlemitsos {\etal} 1995).
In the common focal plane, two Solid-state Imaging Spectrometers
(SIS: Burke {\etal} 1994, Yamashita {\etal} 1997)
and two Gas Imaging Spectrometers
(GIS: Makishima {\etal} 1996, Ohashi {\etal} 1996) are mounted.
The SIS has high sensitivity in the lower energy bandpass
and high energy resolution of $\Delta E/E \simeq 0.02$ (at the time of launch),
whereas the GIS has high throughput in the higher energy bandpass and
high time resolution.

Throughout the observation, the GIS operated in Pulse Height normal mode
in which the band 0.7-10~keV is covered by 1024 pulse height channels.
The SIS mode was switched between the 1-CCD FAINT mode in high bit rate
and the 1-CCD BRIGHT mode in medium bit rate, which cover 0.4-10~keV with
4096 and 2048 pulse height channels, respectively.
The observation was performed normally except for
loss of data during October 1.57--1.83 (UT)
because of a sudden cancelation of the Deep Space Network service.

\begin{figure}[tbh]
   \vbox{\psfig{figure=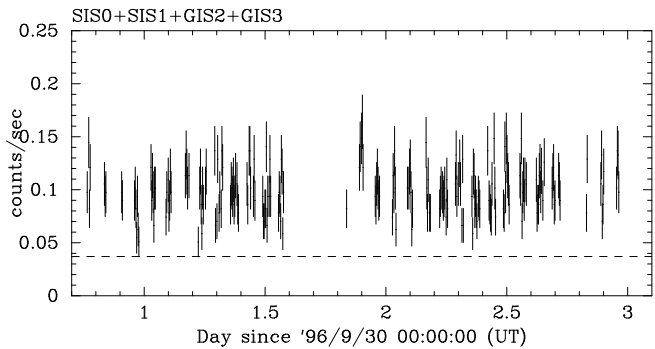,width=8.8cm,%
      bbllx=.4cm,bblly=1.9cm,bburx=15.9cm,bbury=10.0cm,clip=}}\par
   \caption{Light curve of {\rxj} from all the four detectors
	in the band 0.5--10~keV with 256~sec binning.
        The horizontal dashed line indicates the background level.}
   \label{LC}
\end{figure}

\section{Analysis and Results}

\subsection{Data Selection}

We have screened the data with the following criteria.
The data taken while the spacecraft passes the South Atlantic Anomaly
are discarded. In order to avoid the Earth-limb effect,
we have only chosen  data when the Earth elevation angle of {\rxj}
exceeds 5$^\circ$. In addition to this, we have also discarded the SIS data
while the elevation angle from the sunny Earth limb is less than 10$^\circ$.
For the SIS, we have skipped day-night transition periods of the spacecraft
which occur during every satellite orbit.
With these selection criteria, some 77~ksec exposure time is
retained for both the SIS and the GIS.

For the integration of the X-ray source photons,
we have adopted an aperture of 3\farcm7 and 4\farcm0 in radius
centered on {\rxj} for the SIS and GIS, respectively.
For the background, the entire CCD chip outside the aperture is used for
the SIS (there are no other X-ray sources within the field of view),
whereas an annular region which has the same distance from the
boresight of the XRT as the source-integration region is adopted for the GIS.

In Fig.~\ref{LC}, we show the light curve of {\rxj} from all the four
detectors in the 0.5--10~keV band with 256~sec binning after the data screening
as described above.
Because {\asc} is a low-Earth orbit satellite,
the source is usually occulted by the Earth every 96~min. (satellite orbital
period).
As mentioned in \S~2, approximately 6~hrs of data are lost in the middle of
the observation due to a failure of data retrieval on a ground station.
The average background-subtracted counting rate is
0.052~c~s$^{-1}$ with all the four detectors
(0.035~c~s$^{-1}$ for the two SIS, and 0.017~c~s$^{-1}$ for the two GIS).

\begin{figure}[tbh]
      \vbox{\psfig{figure=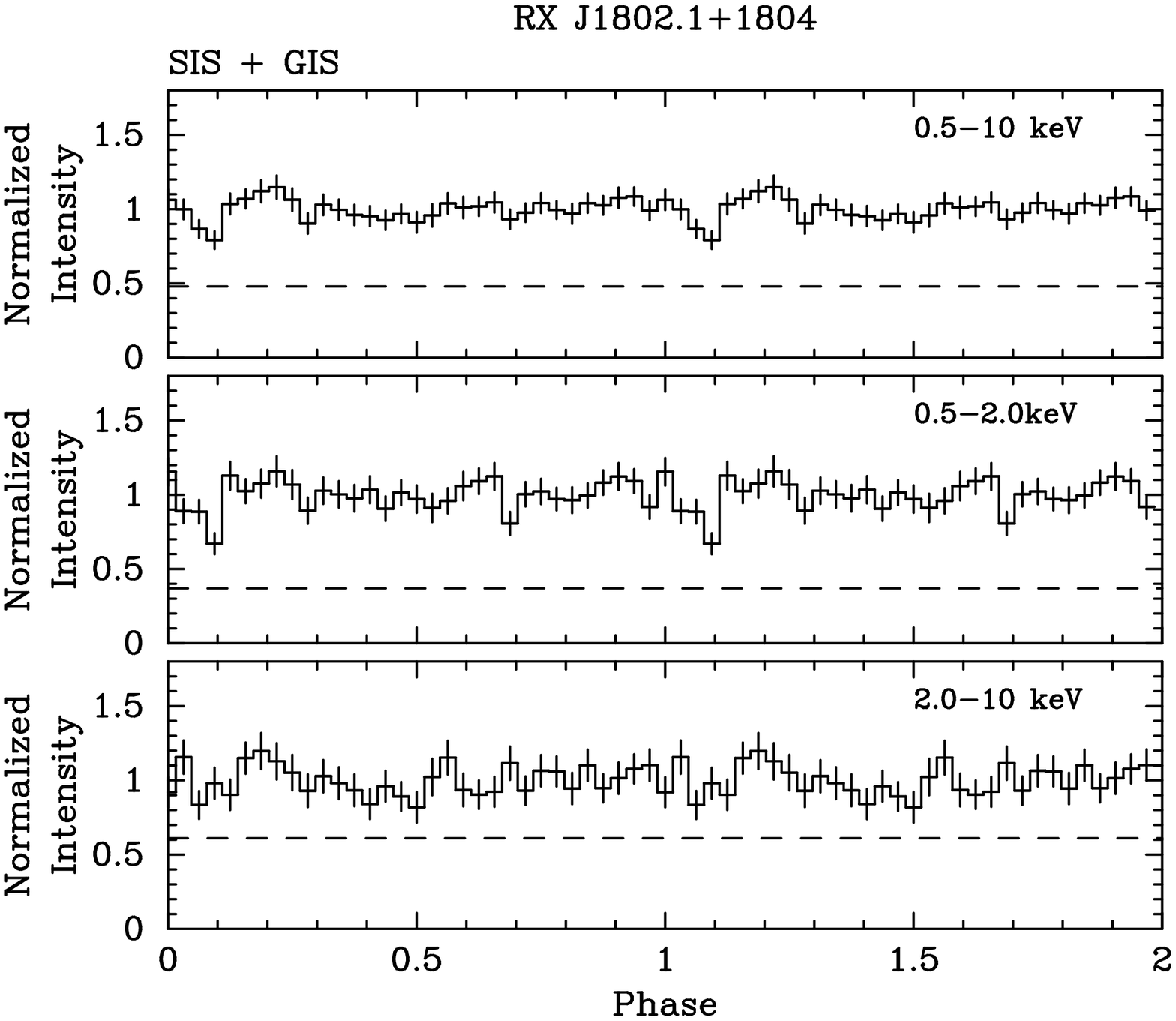,width=8.8cm,%
          bbllx=3.7cm,bblly=3.cm,bburx=23.8cm,bbury=19.4cm,clip=}}\par
   \caption{Energy-resolved light curves of {\rxj} folded with the {\ro} 
   ephemeris (eq.(\ref{ROSephem}), Greiner, Remillard and Motch 1995, 1998).
  The energy bands of the light curves are shown
  at the upper right corner of each panel.
  Dashed lines indicate the background level in each panel.}
\label{FLC}
\end{figure}

\subsection{Energy-resolved Light Curves}

Fig.~\ref{FLC} shows the folded light curves from all the four detectors
in the 0.5--10~keV band and in three separate energy bands.
In folding the light curves, we have adopted the ephe\-me\-ris 
determined from {\ro} and optical observations
(Greiner, Remillard and Motch 1995, 1998), which is

\begin{equation}
T(HJD) = 2449242.3124(21) + 0.07847977(11) \times E.
\label{ROSephem}
\end{equation}

This ephemeris is accurate enough to predict the time of the dip
with an error of only several minutes (phase uncertainty of $\pm 0.02$)
at the time of the {\asc} observation. However, unlike the {\ro} light curves 
which show a modulation amplitude of 100\% below 0.5~keV,
the {\asc} light curves are extremely flat except for a possible dip
during the phase 1.0--1.1. From the average counting rates of the 
two phase bins around  phase 1.08, we have obtained the depth of the dip
to be $33\pm 14$\%, $35\pm 23$\% and $<54$\% of the phase-averaged intensity
for the bands 0.5--10~keV, 0.5--2~keV and 2--10~keV,
respectively, at the 90\% confidence level.
With the dip duration and  depth,
it is possible for this dip to correspond to that found in the {\ro}
light curve, although this is not conclusive.
If we assume that the dip seen in Fig.~\ref{FLC} is 
the same as the one in the {\ro} observation, 
the best-fit period would become slightly longer (0.07848022(16) days)
but inconsistent with the ROSAT folding.

\begin{figure}[thb]
      \vbox{\psfig{figure=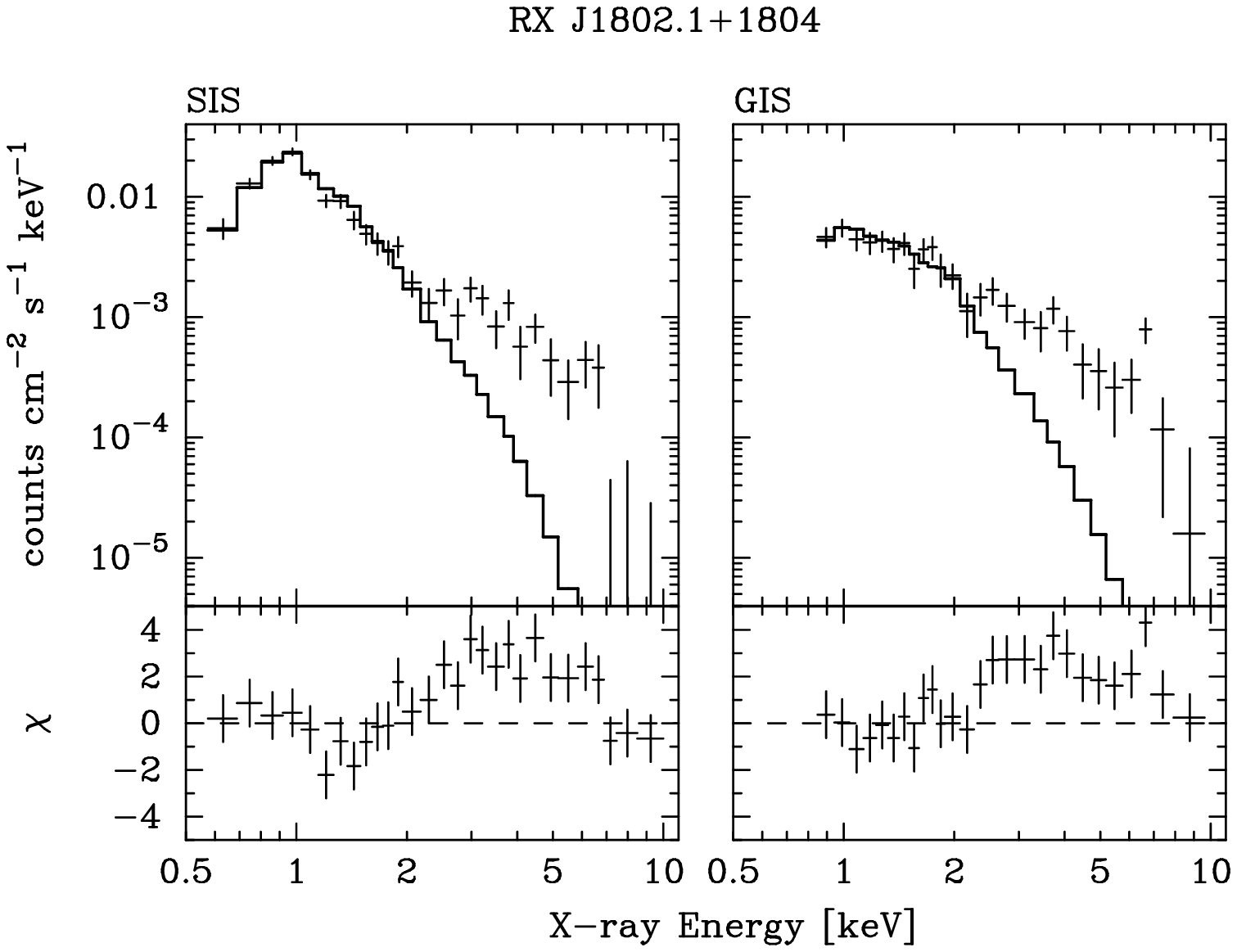,width=8.8cm,%
          bbllx=.2cm,bblly=0.3cm,bburx=15.9cm,bbury=11.3cm,clip=}}\par
     \caption{Combined spectral fit to the SIS and the GIS data with a single 
       Raymond-Smith model.}
     \label{WARA}
      \vbox{\psfig{figure=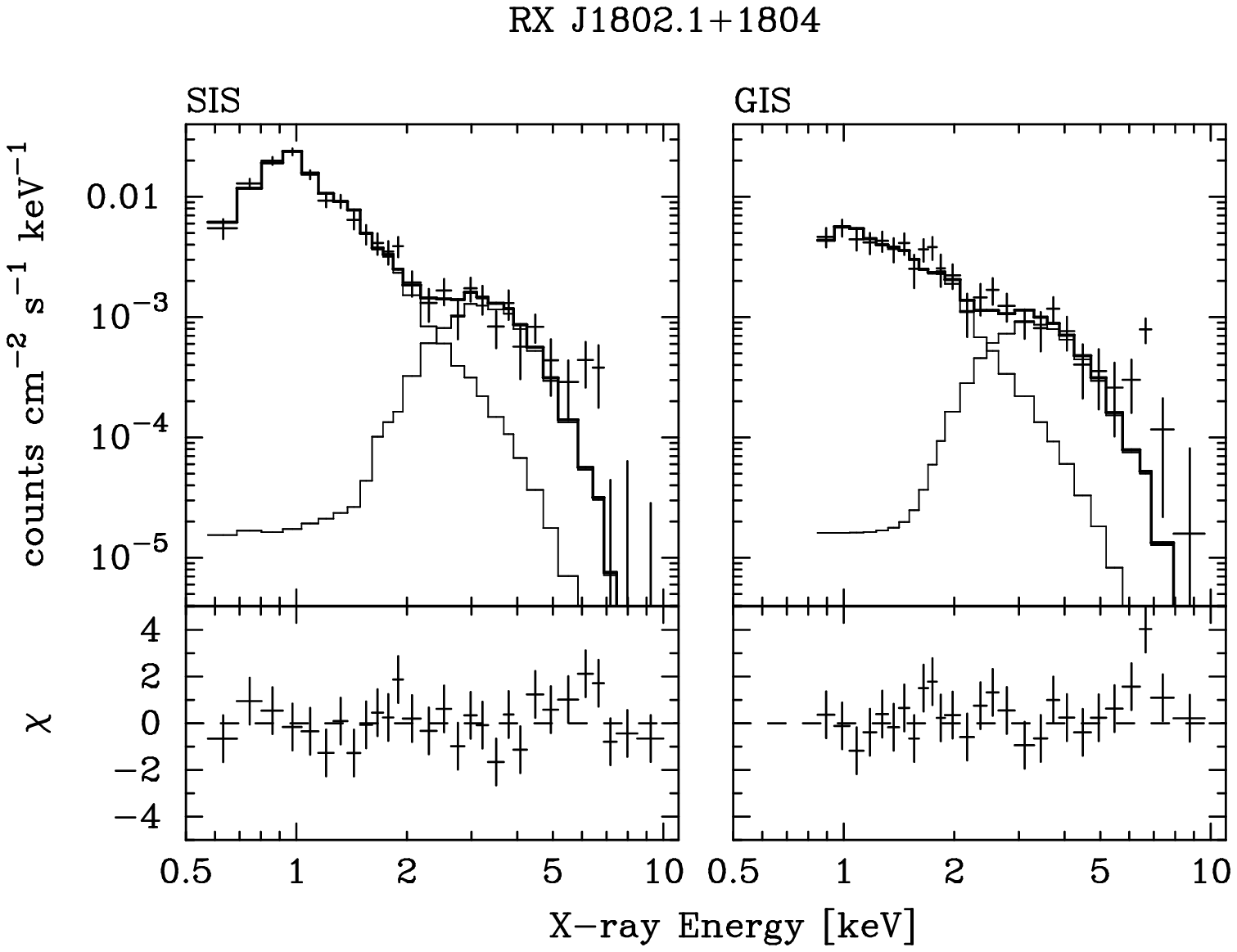,width=8.8cm,%
          bbllx=.2cm,bblly=0.3cm,bburx=15.9cm,bbury=11.3cm,clip=}}\par
     \caption{Combined spectral fit to the SIS and the GIS data with a single 
     Raymond-Smith model undergoing partial covering absorption.}
\label{WARAWARA}
\end{figure}

\subsection{ASCA Spectra}

Since no remarkable intensity variation is found during the observation, 
we sum up all the data.
According to the data selection criteria described in \S~2,
we have extracted the source and the background spectra separately for
the four detectors.
Then we have summed the two SIS spectra and the two GIS spectra,
and have created background subtracted SIS and  GIS spectra, respectively.
Hereafter, we derive spectral parameters of {\rxj} by a combined fit of 
the SIS and GIS spectra using XSPEC version 9.01 (Arnaud 1996).

\begin{table*}[tbh]
\begin{small}
\begin{center}
\caption{Spectral parameters of the combined fits to the SIS and the GIS spectra.}
\label{ASCApara}
\begin{tabular}{llcccccccc}\hline \hline
Model        &                                   & 1Abs$\ast$1RS   & 2Abs$\ast$1RS        & 2Abs$\ast$1RS        & 1Abs$\ast$2RS        & 1Abs$\ast$2RS         \\
             &                                   &                 &                      &  +GA                 &                      &  +GA                  \\ \hline
$kT_1$       & [keV]                             & $\sim$ 0.9      & 0.97 $^{1.03}_{0.84}$& 0.94 $^{1.02}_{0.82}$& 0.86 $^{0.89}_{0.82}$& 0.87 $^{0.89}_{0.84}$ \\
$kT_2$       & [keV]                             & ---             & ---                  & ---                  & 10.3 $^{18.9}_{6.7}$ & 30 $^{\infty}_{7.1}$  \\
Abundance    & [Solar]                           & $\sim$ 0.1      & 0.14 $^{0.23}_{0.09}$& 0.13 $^{0.21}_{0.08}$& 5.8 $^{16.9}_{2.5}$  & 0.82 $^{8.3}_{0.26}$  \\ \hline
$N_{\rm H1}$ & [$10^{21}$ cm$^{-2}$]             & $\sim$ 0.1      & $<1.6$               & $<$1.9               & $<$ 0.6              & $<$ 0.8               \\
$N_{\rm H2}$ & [$10^{21}$ cm$^{-2}$]             & ---             & 130 $^{170}_{100}$   & 130 $^{170}_{100}$   & ---                  & ---                   \\
Cover. Frac. & [\%]                              & ---             & 96 $^{97}_{95}$      & 97 $^{98}_{96}$      & ---                  & ---                   \\ \hline
Line Center  & [keV]                             & ---             & ---                  & 6.55 $^{6.64}_{6.47}$& ---                  & 6.55 $^{6.67}_{6.36}$ \\
Intensity    & [$10^{-5}$ phs s$^{-1}$cm$^{-2}$] & ---             & ---                  & 1.4 $^{1.9}_{1.0}$   & ---                  & 0.79 $^{1.16}_{0.42}$ \\
Equiv. Width & [keV]                             & ---             & ---                  & 12 $^{15}_{8}$       & ---                  & 1.8 $^{2.7}_{1.0}$    \\ \hline
$\chi^2_\nu$ (dof) &                             & 3.72(52)        & 1.21(50)             & 0.68(48)             & 0.74(50)             & 0.69(48)              \\ \hline
\end{tabular}
\end{center}
\end{small}
\end{table*}

We have first tried to fit the spectra with a single temperature optically 
thin thermal plasma model (Raymond and Smith 1977, hereafter referred to as 
R\&S model) undergoing photoelectric absorption. We note that a thermal 
bremsstrahlung model has conventionally been used to fit X-ray spectra of mCVs.
However, optically thin thermal plasma also produces plenty of emission lines
from abundant heavy elements. Among them, the iron $L$-lines appearing 
between 0.7--2~keV cause a significant excess above the continuum 
especially in the case that the plasma temperature is lower than $\sim 3$~keV.
It is very difficult to resolve them from the continuum spectrum even with
the high spectral resolution of the {\asc} SIS.
In addition, there are a few more processes in the optically thin thermal
plasma that produce continuum emission, such as the free-bound transition and
the two-photon decay. Under these circumstances, we cannot estimate the 
continuum parameters if we use the thermal bremsstrahlung model.
Hence we will substitute the thermal brems\-strah\-lung model by a R\&S model 
throughout this paper, except when fitting the spectrum in the band 4--10~keV,
because the continuum in this band is always dominated 
by the thermal bremsstrahlung.

The result of the R\&S model fit is shown in Fig.~\ref{WARA}, and the best 
fit parameters are listed in the third column of Table~\ref{ASCApara}.
Although the fit seems good below $\sim$2~keV,
it obviously shows excess emission above $\sim$2~keV.
The most remarkable structure in the residual is the iron emission line
appearing in the 6--7~keV band. If we evaluate this with a thermal 
bremsstrahlung model plus a Gaussian line in the
4--10~keV band, the equivalent width becomes $4.0\pm1.7$~keV.
Note that {\gin} observations indicated that the equivalent width was in the 
range 0.2--0.8~keV for a dozen of mCVs (Ishida and Fujimoto 1995).
Hence the equivalent width of {\rxj} is roughly an order of magnitude larger
than that of mCVs observed by {\gin}, 
indicating a greater abundance by the same order.

In order to explain the excess emission above the single component model,
we have tried the two possibilities described in \S~1, i.e. applying a 
partial-covering absorption model and introducing a second R\&S component.

\begin{figure}[tbh]
      \vbox{\psfig{figure=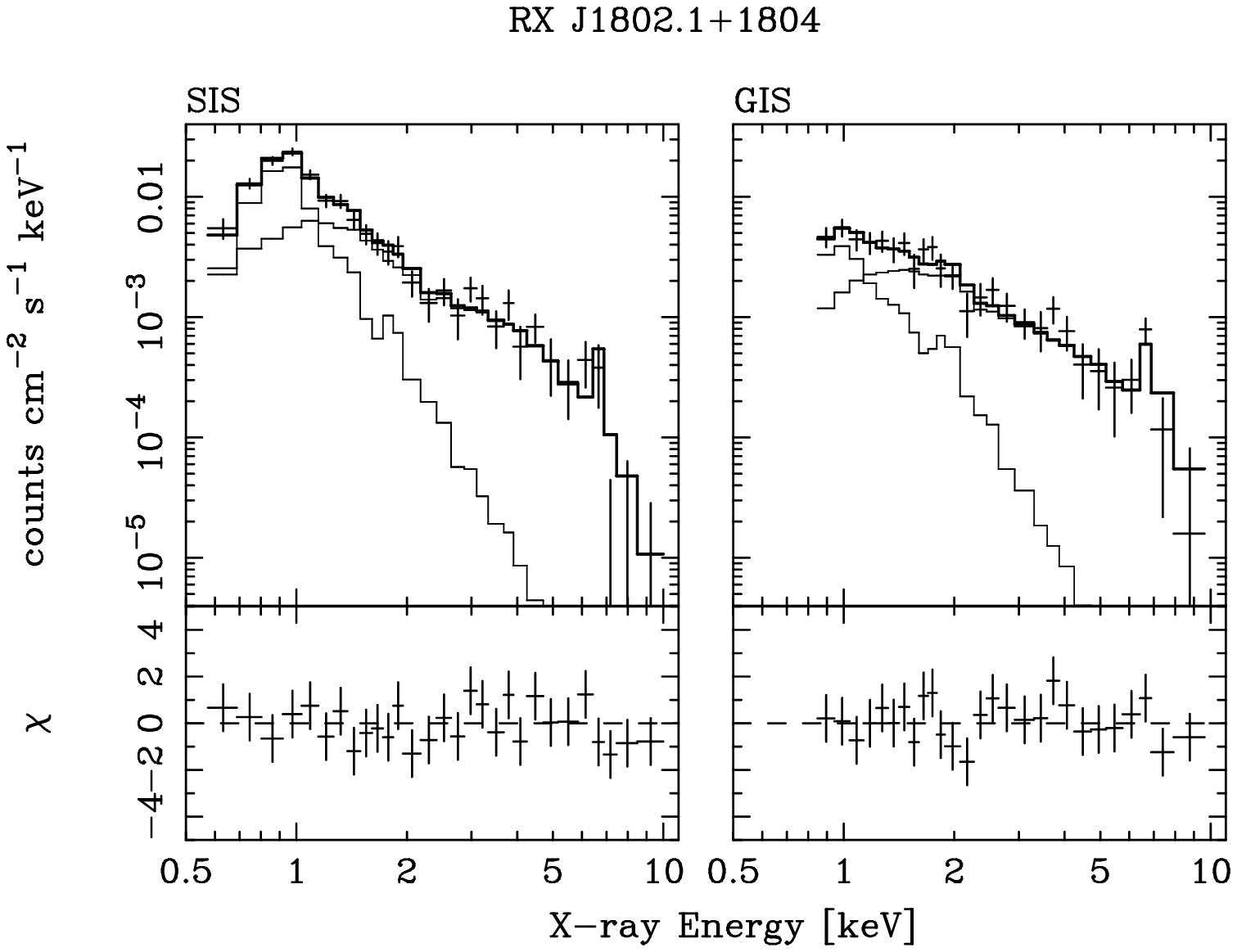,width=8.8cm,%
          bbllx=.2cm,bblly=0.3cm,bburx=15.9cm,bbury=11.3cm,clip=}}\par
    \caption{Combined spectral fit to the SIS and the GIS data with a two 
        temperature Raymond-Smith model.}
    \label{WARARAGA}
\end{figure}

We first have attempted to apply the partial-covering absorption model
to the observed spectra.
The result is shown in Fig.~\ref{WARAWARA} and the best fit parameters are 
summarized in the 4th column of Table~\ref{ASCApara}.
The reduced $\chi^2$ value of 1.21 means that despite the improvement
over the single R\&S model the partial-covering absorption model is only 
marginally acceptable.
The best fit values of the two hydrogen column densities are
$<2\times 10^{21}$ cm$^{-2}$ and $1.3\times 10^{23}$ cm$^{-2}$
and the covering fraction of the latter over the emission region
is $97\pm1$\%. From {\ro} spectra a ratio of soft blackbody flux
to hard bremsstrahlung flux  of nearly 90 in the
0.1--2.4~keV band was deduced (Greiner, Remillard and Motch 1995, 1998).
This extreme soft excess now vanishes 
because of the high covering fraction of the heavily absorbed component.
But the model still does not reproduce the prominent iron emission line
 which is seen between 6-7~keV.
We have thus added a Gaussian,  and have fitted the spectra again.
The result is summarized in the 5th column of Table~\ref{ASCApara} and
shows that the $\chi^2$ value decreases by nearly 30
after adding two free parameters into the model.
Hence, the introduction of the Gaussian is statistically justified.
Note, however, that the resulting line equivalent width becomes $\sim$~12~keV
which is unacceptably large.
Since the temperature of the emission component is lower than 1~keV,
it seems unlikely that this emission line comes from the hot plasma itself.
The fluorescent iron emission line is, on the other hand, expected to
emanate from the white dwarf surface illuminated by the hard X-ray emission.
However, its equivalent width is estimated to be $\sim$~140~eV
(George and Fabian 1991, Done {\etal} 1994, Beardmore {\etal} 1995)
if the white dwarf surface has solar composition of heavy elements.
Therefore, the equivalent width determined from the fit indicates
an abundance of the order of $\sim$100 times Solar,
which is in strong contradiction to the abundance from the R\&S model, 
$\sim$~0.1 (Table~\ref{ASCApara}).
We conclude that the partial-covering absorption model cannot reproduce
the observed spectrum in a physically consistent manner.

\begin{figure*}[tbh]
\begin{minipage}{0.55\textwidth}
      \vbox{\psfig{figure=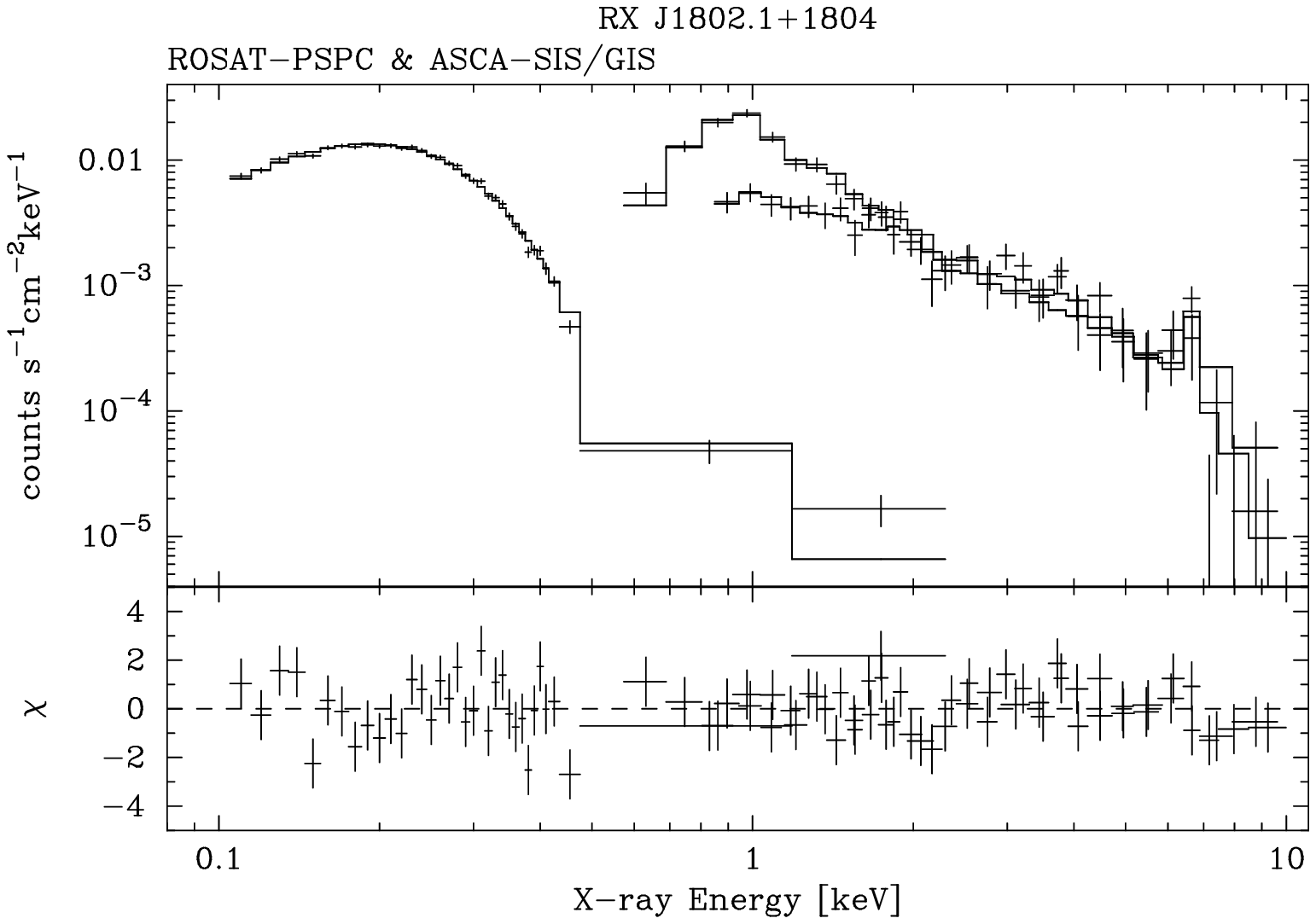,width=10.cm,%
          bbllx=.5cm,bblly=0.3cm,bburx=16.3cm,bbury=11.1cm,clip=}}\par
\end{minipage}
\hfill
\begin{minipage}{0.43\textwidth}
     \vbox{\psfig{figure=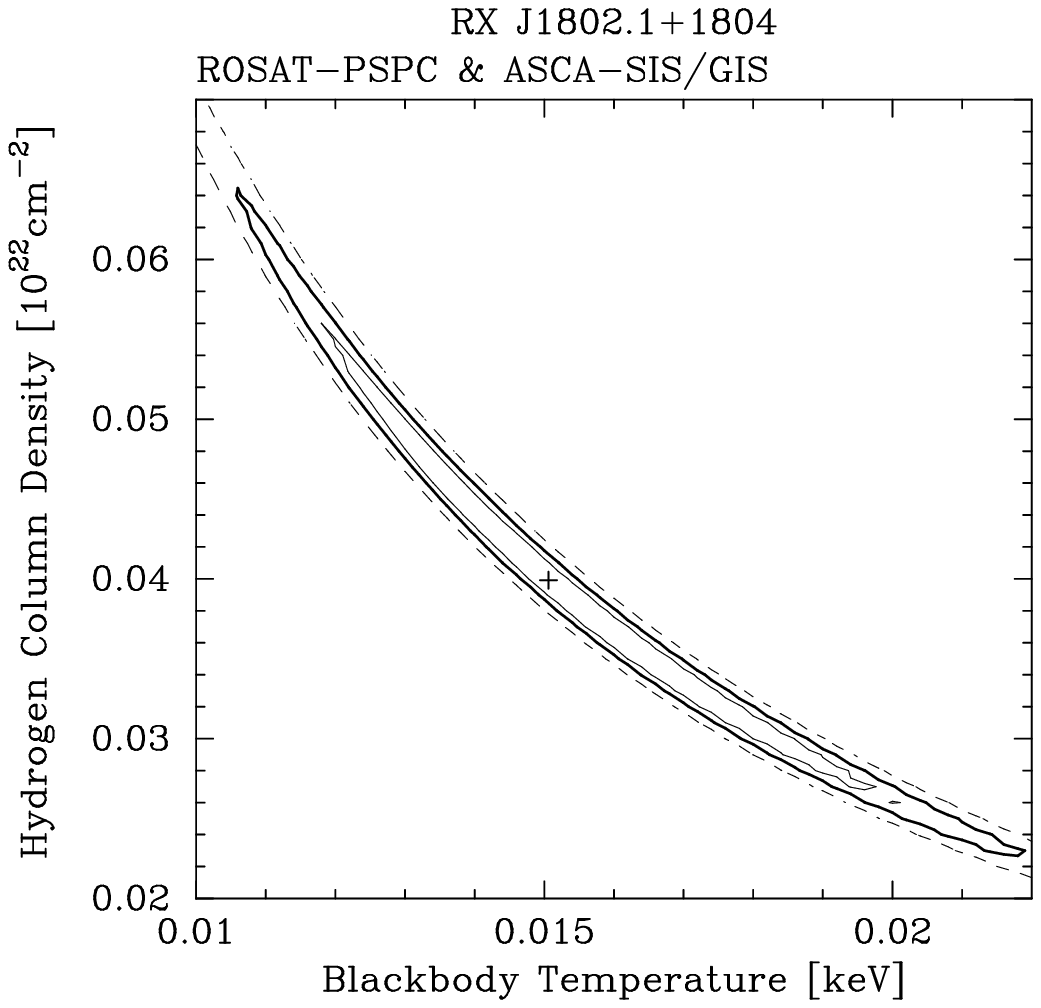,width=7.4cm,%
          bbllx=2.2cm,bblly=0.8cm,bburx=13.0cm,bbury=10.6cm,clip=}}\par
\end{minipage}
 \caption{Combined fit to the {\ro} PSPC and the {\asc} SIS/GIS spectra
 with a blackbody and a two temperature thin thermal plasma model (left),
 and the confidence contours of $N_{\rm H}$ vs. $kT$ for the
 blackbody component. The contours are 68\% (innermost), 90\% and 99\% 
 confidence levels.}
 \label{WABBRARA}
\end{figure*}

As the next step, we have tried to fit the hard excess component shown in 
Fig.~\ref{LC} by introducing a second R\&S component.
The result of the fit is shown in Fig.~\ref{WARARAGA}, and the best fit 
parameters are shown in the 6th column of Table~\ref{ASCApara}.
The fit is acceptable with a reduced $\chi^2$ value of 0.74, suggesting that
the X-ray spectrum of {\rxj} consists of multi-temperature
optically thin thermal plasma emission components.
The obtained flux is 4.8$\times 10^{-13}$ erg s$^{-1}$ cm$^{-2}$
in the band 2--10~keV.

Note that this fit still suggests a very high abundance of 6 times Solar
with a lower limit of 2.5 times Solar, which seems too high for
cataclysmic variables, because they are generally considered to be old systems.

Recently, Hellier \etal\ (1998) compiled spectra of 15
mCVs from {\asc} archival data. A total of 14 spectra out of the 15 show 
a significant fluorescent iron {\ka} emission line at 6.41~keV,
as well as the two thermal plasma components at 6.68~keV and 6.97~keV.
Among them, the fluorescent component probably originates from the
white dwarf surface (Done, Osborne and Beardmore 1995).
Although the statistics of our data is not good enough to resolve these three
components, it is necessary to include the fluorescent iron {\ka} line
into the model in evaluating the abundance correctly.

We thus have introduced a Gaussian line as representing
the iron {\ka} line of fluorescence origin.
The result is summarized in the last column of Table~\ref{ASCApara}.
Although the best fit abundance is reduced to $\sim$0.9,
the equivalent width of the fluorescent iron {\ka} line becomes around 2~keV.
This value again indicates an abundance of more than 10 times Solar.
Clearly, the abundances estimated from the intensities of iron {\ka} lines
of the hot plasma origin and of the fluorescence origin
should be consistent. This point will be discussed in \S~4.
Note also that the high abundance can affect the estimation of the
bolometric luminosity of the hard component,
since the line emission predominates among all the cooling processes in the
plasma the temperature of which is less than 2~keV (McCray 1987).
We therefore calculate the bolometric luminosity of the hard optically thin
thermal component later in relation with the abundance.

\subsection{Combined Spectral Fit of ROSAT and ASCA}

Greiner, Remillard and Motch (1995, 1998) reported that the flux of the
soft blackbody component is greater than that of the hard thin thermal
plasma emission by two orders of magnitude in the band 0.1--2.4~keV.
We have attempted to re-examine this extreme soft excess
in combination with the {\asc} hard X-ray data.

{\ro} pointed {\rxj} four times between 1992 October and 1993 September.
We have extracted a mean {\ro} PSPC spectrum from the observation on 1993
September 11/12 (the exposure time of which was $\sim$~13~ksec, the longest of
all the pointing observations). Details of the observations are presented
in Greiner, Remillard and Motch (1998) (see also Greiner, Remillard and Motch 
1995).

\begin{figure*}[tbh]
     \vbox{\psfig{figure=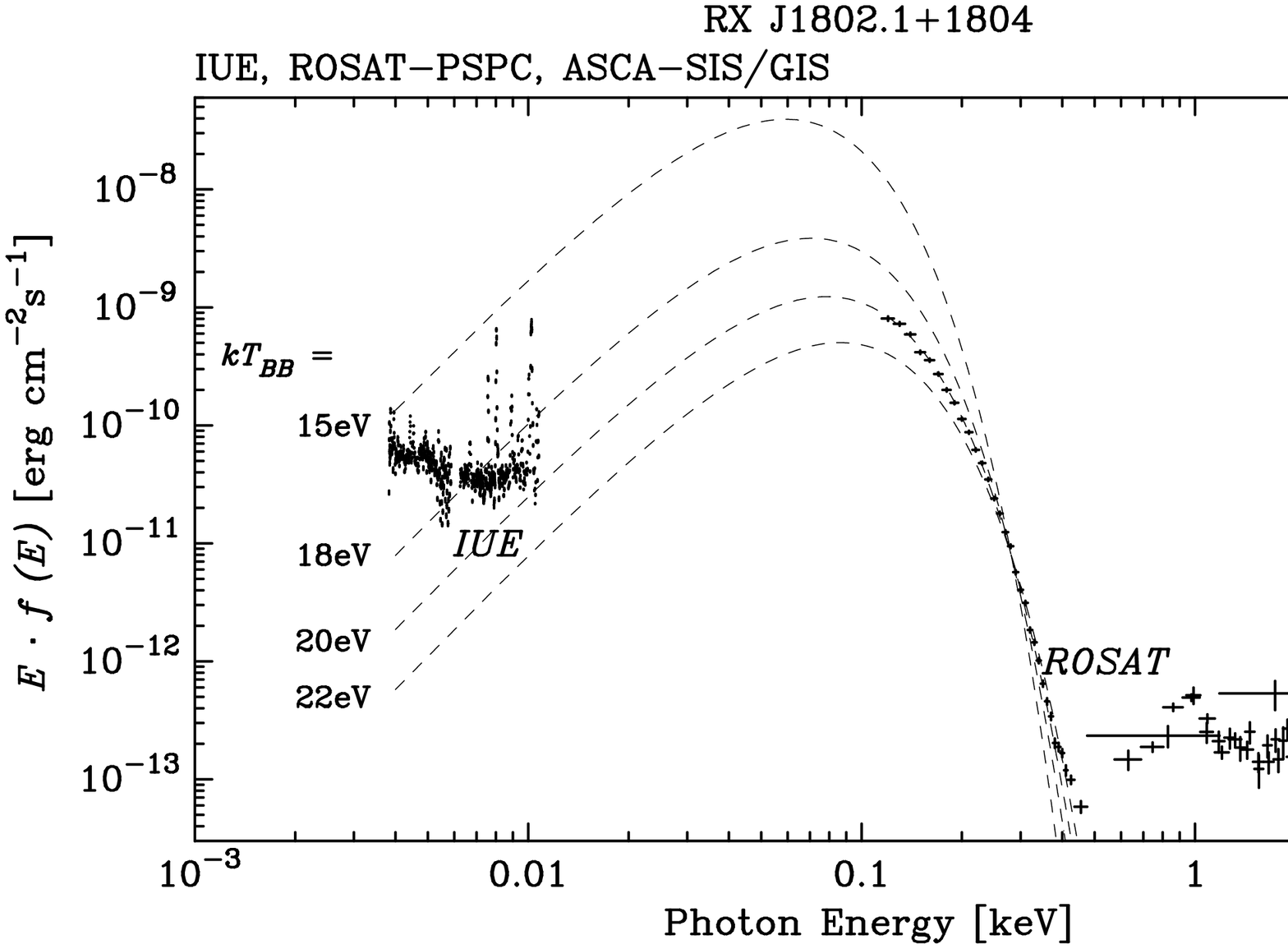,width=11.5cm,%
       bbllx=1.8cm,bblly=4.1cm,bburx=25.6cm,bbury=18.cm,clip=}}\par
   \caption{Intrinsic spectra of {\iue}, {\ro} and {\asc}.
    $N_{\rm H}$ is corrected for the latter two.
   The temperature of the blackbody is assumed to be 20~eV for the
   {\ro} spectrum (\S\S~3.4).
   For the {\iue} spectrum, extinction correction is smaller
   than a factor 2 and has not been applied.
   $N_{\rm H}$- corrected blackbody models from 
   the {\ro} and {\asc} combined fitting are also drawn by dashed
   lines in the temperature range 15--22~eV.
   The allowed range of the blackbody temperature becomes tightly constrained
   to 20-22~eV by including the {\iue} data (Shrader {\etal} 1997).}
   \label{AllCombine}
\end{figure*}

Since the {\ro} observation is not simultaneous with the {\asc} observation,
we have first checked if the intensity levels of the two observations
are consistent. To do this, we have used the {\ro} PSPC and the {\asc} SIS 
and GIS spectral channels below 2~keV, and have fitted a model consisting of 
a soft blackbody and a hard thin thermal plasma spectrum
undergoing photoelectric absorption represented by a common hydrogen
column density. Although the temperatures of both components, 
the normalization of the blackbody as well as
the abundance of the thin thermal plasma are constrained
to be the same among the three spectra,
the normalization of the thin thermal component is set free to
vary independently (note that the blackbody parameters are determined solely
by the PSPC spectrum).
The resulting normalizations of the hard thin thermal emission of the
SIS/GIS relative to that of the PSPC are 1.03 and 1.09, respectively,
with a typical statistical error of $\sim \pm 0.3$.
We thus regard the intensity level of the hard component as the
same between the {\ro} and the {\asc} observations.

Next, we have performed a combined spectral fit in the entire 0.1--10~keV band
with a model composed of a blackbody and a two temperature thin thermal
plasma emission component. The result is shown in Fig.~\ref{WABBRARA}.

The fit is marginally acceptable at the 90\% confidence level,
with $\chi^{2}_{\nu}$ of 1.15 for 83 degrees of freedom.
The confidence contours for the hydrogen column density and the temperature
of the blackbody component are also shown in Fig.~\ref{WABBRARA}.
The best fit temperature of the blackbody is obtained to be
$15^{+7}_{-5}$~eV for  two parameters of interest.
The observed flux of the blackbody is
$5.6\times 10^{-12}$ erg cm$^{-2}$ s$^{-1}$ in the band 0.1--2.4~keV.
After $N_{\rm H}$ correction, the flux becomes 
$5.1\times 10^{-9}$ erg cm$^{-2}$ s$^{-1}$, which 
is $\sim 10^4$ times as large as that of the hard component in the 2--10~keV
band (\S\S~3.3). The major difference to Greiner, Remillard and Motch (1998) is
that the blackbody temperature now is even lower than in 
the fit of the \ro\ data alone, because a part of the emission around 
$\approx$0.5 keV is now attributed to the low-kT R\&S component.

Because the blackbody temperature is very low,
it is very difficult to calculate the bolometric luminosity of the blackbody,
because even the {\ro} PSPC can observe only the high energy end of the Wien
tail. Assuming a disc-shape emission region and a distance of 100~pc,
we obtain a bolometric luminosity of the blackbody component
$L_{\rm BB}$ of $1.6\times 10^{34}$ erg s$^{-1}$/$<\cos \theta>$
for the best fit temperature of 15~eV,
where $<\cos \theta>$ implies the cosine of the angle between the normal
of the disk and the line of sight.
However, the 90\% confidence range of $L_{\rm BB}$ becomes
$2\times 10^{32} - 1\times 10^{37}$ erg s$^{-1}$
for the temperature of 22--10~eV.

\subsection{Constraint on the Soft Component from IUE Data}

To further constrain the spectral parameters of the soft (blackbody) component,
we have utilized the {\iue} observation performed on 1995 Aug.~31 
(Shrader {\etal} 1997).
Among the several exposures of {\iue}, we collected the data taken
out of the X-ray eclipse, which are SWP55775/6 and LWP31382 using the
ephemeris of Greiner, Remillard and Motch (1998).
For SWP, we took the time average of the two exposures.
In Fig.~\ref{AllCombine}, we have plotted the {\iue} spectra thus obtained
together with $N_{\rm H}$-corrected spectra of {\ro} ($kT=20$~eV) and {\asc}
in a $\nu F(\nu)$ diagram.
Note that we have not corrected the color of the {\iue} spectrum.
The correction factor at 1000~{\AA} is, however, less than 2
from the hydrogen column density $3.2\times 10^{20}$~cm$^{-2}$
(Greiner, Remillard and Motch 1998), and even smaller at longer wavelengths.

We have also shown the $N_{\rm H}$-corrected best fit blackbody spectra
obtained from the simultaneous {\ro} and {\asc} spectral fitting (\S\S~3.4).
It is obvious that the shape of the {\iue} spectrum does not 
correspond (at least in the long-wavelength region)
to that of the blackbody extrapolation, and
therefore represents a separate emission component. Thus, 
the allowed $N_H$-corrected blackbody curves have to be below the {\iue} 
spectrum, constraining the blackbody temperature  to the range
20--22~eV. The corresponding bolometric luminosity of the blackbody
is obtained to be 2--5$\times 10^{32}$ erg s$^{-1}$/$<\cos \theta>$
for a distance of 100~pc.
We note that using a white dwarf atmosphere model will predominantly reduce 
$L_{\rm BB}$ but result in a rather similar effective temperature.

\section{Discussion}

\subsection{Possible Accretion Geometry}

As shown in \S\S~3.2, the {\asc} folded light curve shows little evidence of
orbital modulation, while the low energy {\ro} light curves show 
a deep X-ray intensity modulation with an  amplitude of 100\%.
This probably implies that the accretion pole moves around in the hemisphere
of the white dwarf which is visible from the observer,
and the X-ray modulation is caused by photoelectric absorption
in the accretion column,
as is also the case for the recently discovered new polar {\axj}
(Misaki {\etal} 1996, Thomas and Reinsch 1996).

From the {\ro} and {\asc} light curves, we can estimate the hydrogen column
density of the accretion column passing over the line of sight.
Assuming the absorption by the column can be represented by
a single hydrogen column density and the absorbing matter is cold, 
the PSPC response function predicts $N_{\rm H}>8\times 10^{20}$~cm$^{-2}$
for the {\ro} counting rate in the band 0.1--0.5~keV to be reduced by 95\%,
if the blackbody parameters are the same as those out of the eclipse.
On the other hand, for the {\asc} counting rate (SIS+GIS) to be
reduced less than 20\% in the band 0.5--10~keV,
the SIS and the GIS responses require $N_{\rm H}<2\times 10^{21}$~cm$^{-2}$.
Therefore, the hydrogen column density of the accretion column passing
over the line of sight at the time of eclipse is $\sim 10^{21}$~cm$^{-2}$
on condition that the absorber is cold and can be characterized by a single
hydrogen column density.
As noted in Greiner, Remillard, Motch (1998), however,
a detailed analysis of the energy resolved light curve of {\ro}
indicates that the absorption by the column can hardly be reconciled with
a single hydrogen column density.
It is possible that the pre-shock column is ionized in part or has a
distribution in $N_{\rm H}$ in the range \lax 10$^{21}$~cm$^{-2}$.

\subsection{Evidence of Postshock Cooling Flow}

As explained in \S~1, the hard X-ray spectrum of mCVs can usually be
modelled by thermal bremsstrahlung with a single temperature
in the range 10--40~keV (Ishida and Fujimoto 1995).
Although theories of the postshock accretion flow predict that
the postshock plasma is cooled via thermal bremsstrahlung (Aizu 1973)
and also cyclotron radiation (Wu {\etal} 1994, Woelk and Beuermann 1996)
evidence of this cooling has been difficult to find observationally,
because a thick ($N_{\rm H}\sim 10^{23}$ cm$^{-2}$),
partial-covering absorption caused by the accretion column
prevents us from measuring the shape of the intrinsic spectrum.

The only exception is {\ex} in which the hard X-ray continuum emission
can be represented by a two temperature R\&S model ($kT=$~0.8~keV and 8~keV)
at first order approximation,
and the ionization temperatures of heavy elements distribute in the
range 0.9--8~keV (Ishida, Mukai and Osborne 1994).
Fujimoto and Ishida (1997) showed that 
the distribution of the ionization temperatures is consistent with
the postshock cooling flow predicted by Aizu (1973), and successfully 
determined the shock temperature and the mass of the white dwarf.
The X-ray spectrum of {\rxj} also requires two temperature R\&S components,
and is similar to that of {\ex}.
We believe that {\asc} observed the postshock cooling flow in {\rxj}.

This finding probably indicates that the temperature distribution
due to  the postshock cooling is a common feature among mCVs,
and one always finds this as long as the low energy absorption is weak enough
($\leq 10^{21}$~cm$^{-2}$) as in {\ex} and {\rxj}.

\subsection{Abundance}

As displayed in \S~3, the hard part of the X-ray spectrum of {\rxj} has a 
strong iron {\ka} emission line with an equivalent width of $\sim$~4~keV.
The line originates from the hot plasma, and also, probably from the
white dwarf surface via fluorescence (Hellier, Mukai and Osborne 1998).
However, these two components cannot be resolved because of statistical 
limitations. In obtaining the elemental abundance of the plasma, we have to mix
these two components so that the abundances they give are consistent.

\begin{figure}[tbh]
      \vbox{\psfig{figure=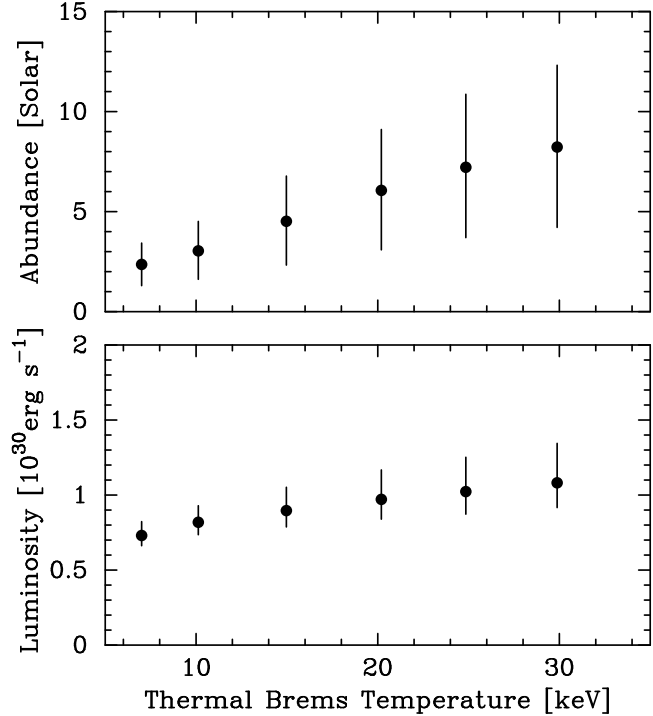,width=8.8cm,%
          bbllx=2.9cm,bblly=1.3cm,bburx=12.8cm,bbury=12.2cm,clip=}}\par
     \caption{The iron abundance of the plasma (upper panel) and 
      the bolometric luminosity (lower panel) of the hard X-ray emission.
      For the luminosity, the distance to the source is assumed to be 100~pc.}
      \label{AbLumi}
\end{figure}

As shown in Table~\ref{ASCApara}, the temperature of the hard excess component
is uncertain, with a lower limit of $\sim$~7~keV.
Therefore, we have fixed the temperature of the plasma at several trial values
between 7 and 30 keV, and have made the following analysis.
First, we have adopted  thermal bremsstrahlung as the continuum spectrum.
Then we have added three Gaussian lines which represent the fluorescent
component at $\sim$~6.4~keV, and He-like and hydrogenic components
at 6.68~keV and 6.97~keV, respectively, 
and have performed the spectral fitting in the band 4--10~keV.
In doing this, we have assumed that all the lines are narrow.
Also the line central energies of the plasma components are fixed at
6.68~keV and 6.97~keV.
The intensities of all the lines are constrained so that they give the same
abundance at each fixed temperature.
As mentioned in \S\S~3.3, the equivalent width of the fluorescent component
should be 140~eV, almost irrespective of the plasma temperature, 
if the white dwarf surface has Solar composition.
On the other hand, the equivalent widths of the plasma components
can be obtained from the atomic data table in Raymond and Smith (1977)
or Mewe {\etal} (1985) as a function of the plasma temperature
in the case of Solar composition plasma.
Therefore, the free parameters of the lines are only two --- 
the central energy and the normalization of the fluorescent {\ka} line.
The fitting result thus obtained is shown in Fig.~\ref{AbLumi}.

The resulting abundance is larger for higher trial temperature.
This is a result of the fact that $K-$shell electrons are increasingly 
stripped off for higher
temperatures, and hence a higher iron abundance is necessary to account
for the observed equivalent width.
The smallest abundance is obtained to be 2.4$\pm 1.1\odot$ at a
temperature of 7~keV, which implies the lower limit of the abundance 
to be 1.3$\odot$.
Note, however, that this is a very conservative lower limit, and 
the abundance based on the iron {\ka} emission line
is probably several times as large as that of Solar composition.
This is in contrast to the abundances of CVs 
which have recently been measured to be sub-Solar, such as 
$0.63\pm 0.08\odot$ for {\ex} (Fujimoto and Ishida 1997), 
$0.4^{+0.2}_{-0.1}\odot$ for {\am} (Ishida {\etal} 1997),
and $\sim 0.4\odot$ for SS~Cyg (Done and Osborne 1997).
A hint for  a larger abundance than Solar
is obtained only for {\axj} (Misaki {\etal} 1996).

\subsection{Luminosity of the Hard X-ray Component and the Soft Excess}

Since we have obtained the abundance in the previous section,
we have next calculated the bolometric luminosity of the hard X-ray
component.
To do this, we have adopted the volume emissivity formulas
of the optically thin plasma approximated by McCray (1987), but
modified to take into account the abundance effects.
\[
\Lambda(T, Z/Z_\odot)\; =\, 1.0\times 10^{-22}
		\left( \frac{Z}{Z_\odot}
			\right) T_6^{-0.7}\; +\]
\[\hspace{2.4cm} 
2.3\times 10^{-24}T_6^{0.5}\hspace{1em} \mbox{[erg cm$^3$ s$^{-1}$]},
\]
where $T_6$ is the plasma temperature in $10^6$~K.
The first term on the right hand side is the volume emissivity for the line
emission which is proportional to the abundance.
The second term represents that of the free-free emission.
Note that the first term is greater than the second term in the range 
$T < 2$~keV.
The bolometric luminosity of the hard component $L_{\rm H}$ is obtained by 
$\Lambda \cdot EM$,
where $EM$ is the emission measure obtained from the spectral
fitting for the 0.8~keV component and the hard excess component separately
by assuming a distance to the source.
The luminosity thus calculated for the trial temperatures is plotted
in the lower panel of Fig.~\ref{AbLumi} showing a rather flat dependence
with temperature in the 7--30~keV range:  
$L_{\rm H}=0.6-1.4\times 10^{30}$ erg s$^{-1}$.
Note that we have not corrected for reflection from the white dwarf surface.
One can do this by dividing the above value by $1+a_X$ where $a_X$ is the
hard X-ray albedo.

In \S\S~3.4, we have obtained the lower limit of the bolometric luminosity
of the blackbody component $L_{BB}$ to be $2\times 10^{32}$ erg s$^{-1}$ from 
{\ro} and {\asc} simultaneous spectral fitting.
This means $L_{\rm S}/L_{\rm H} > 140/<\cos \theta> $.
If we also take {\iue} data into account (\S\S~3.5),
$L_{BB}$ is constrained in the range $2-5\times 10^{32}$ erg s$^{-1}$,
and hence $L_{\rm S}/L_{\rm H} = (140-830)/<\cos \theta>$ is obtained.
Note that white dwarf atmosphere models could possibly reduce the luminosity 
of the soft component, and thus also $L_{\rm S}/L_{\rm H}$.

\subsection{Note on Determining Parameters of the Blackbody Spectrum}

In \S\S~3.4, we have derived the temperature of the soft blackbody component
to be $15^{+7}_{-5}$~eV.
The best fit value is outside the `usual' range derived by
Szkody {\etal} (1995), namely 20--45~eV.
In estimating the blackbody temperature, Szkody {\etal} (1995) assumed 
a thermal brems\-strah\-lung component  with a temperature of 10~keV
for the hard X-ray component.
However, based on our {\asc} data we have found a spectral component 
which can be represented by a R\&S spectrum with $kT \sim$~0.8~keV.
The R\&S component with such low temperature has a forest of iron emission 
lines in the 0.8--1~keV band caused by the iron L-shell transitions
(Raymond and Smith 1977).
Hence, a significant amount of the flux in the 0.8--2~keV band
is attributed to the low temperature R\&S component in our modelling.
Note that this cannot happen if we assume a thermal bremsstrahlung component
with a temperature of 10~keV.
As a result, the blackbody temperature becomes lower than
the estimates in Szkody {\etal} (1995).

In analyzing {\ro} data, one usually assumes the temperature of the hard
X-ray component to be around 20~keV  (Ramsay {\etal} 1994, for example).
As shown here, however, this may cause a huge systematic error
in evaluating the luminosity and the temperature of the soft blackbody
component.

\section{Conclusion}

We presented X-ray data of {\rxj} obtained by {\asc}. From the {\asc} light 
curves we find only marginal evidence for 
orbital intensity modulation which is seen in the {\ro} light curve
below 0.5~keV characterized by the sharp and deep minima. From this energy 
dependence, we conclude that the intensity modulation is caused mostly by
photoelectric absorption in the pre-shock accretion column, 
and the accreting pole moves around on the hemisphere
visible from the observer, consistent with the conclusions from
Greiner, Remillard \& Motch (1998).
It is possible that 
the line of sight absorber is partly ionized or has a distribution in
$N_H$ in the range \lax 10$^{21}$~cm$^{-2}$.

The X-ray spectrum can be represented by a two temperature
optically thin thermal plasma emission model with temperatures of
$\sim$~1~keV and $>$~7~keV. In analogy with {\ex}, 
we deduce that {\asc} observed the cooling of the postshock plasma,
as indicated by the theory of the postshock accretion flow.
A remarkable feature of the X-ray spectrum of \rxj\ is the strong iron 
{\ka} emission line whose equivalent width is $\sim 4$~keV.
To account for this, an iron abundance greater than Solar by
at least 1.3 times is required. 
From the combined analysis of the {\ro} PSPC and {\iue} spectra,
the ratio of the bolometric luminosity of the soft component to the hard
is revealed to be greater than 140.

\begin{acknowledgements}
We are grateful for Dr. C.R. Shrader for supplying us with his {\iue} spectra.
MI greatly appreciates financial support from JSPS along the
Japan-Germany collaboration programme.
JG is supported by the Deutsche Agentur f\"ur Raumfahrtangelegenheiten
(DARA) GmbH under contract numbers FKZ 50 OR 9201 and 50 QQ 9602\,3.
RR acknowledges partial support from NASA grant NAG5--1784.  
\end{acknowledgements}


\begin{thebibliography}{}

\bibitem{a73} Aizu K., 1973, Prog. Theoret. Phys. {49}, 1184

\bibitem{arn96} Arnaud K.A., 1996, Astronomical Data Analysis Software and
           Systems V, eds. Jacoby G. and Barnes J., ASP Conf. Ser. 101, p. 17

\bibitem{bdo95} Beardmore A.P., Done C., Osborne J.P. and Ishida M., 1995, 
{\mnras} {272}, 749

\bibitem{bb95} Beuermann K. and Burwitz V., 1995, 
in Cape Workshop on Magnetic Cataclysmic Variables, eds. D.~A.~H.~Buckley and
B.~Warner (ASP; San Francisco), p.~99

\bibitem{bet94} Burke B.E. {\etal}, 1994, {IEEE Trans. Nucl. Sci.} {41}, 375

\bibitem{c90} Cropper M., 1990, Sp.Sci.Rev. {54}, 195
\bibitem{dmm92} Done C., Mulchaey J.S., Mushotzky R.F. and Arnaud K.A., 1992,
	{\apj} {395}, 275
\bibitem{dob95} Done C., Osborne J. P. and Beardmore A. P. 1995, {\mnras}
	{276}, 483
\bibitem{do97} Done C. and Osborne J. P., 1997, {\mnras} {288}, 649
\bibitem{fkl83} Frank J., King A.R. and Lasota J.P., 1983, {\mnras} {202}, 183
\bibitem{fi97} Fujimoto R. and Ishida M., 1997, {\apj} {474}, 774
\bibitem{gf91} George I.M. and Fabian A.C., 1991, {\mnras} {249}, 352
\bibitem{grm97} Greiner J., Remillard R.A. and Motch C., 1998, A\&A 235, 
   {(this volume)}
\bibitem{grm95} Greiner J., Remillard R.A. and Motch C., 1995,
in Cataclysmic Variables, eds. A. Bianchini,
M. Della Valle and M. Orio (Dordrecht: Holland), p.~161
\bibitem{hmo98} Hellier C., Mukai K. and Osborne J. P. 1998, {\mnras}, 
	{\it in press}
\bibitem{id83} Imamura J. and Durisen, 1983, {\apj} {268}, 291
\bibitem{imo94} Ishida M., Mukai K. and Osborne J. P. 1994, {\pasj} {46}, L81
\bibitem{if95} Ishida M. and Fujimoto R., 1995, 
in Cataclysmic Variables, eds. A. Bianchini,
M. Della Valle and M. Orio (Dordrecht: Holland), p.~93
\bibitem{imfmo97} Ishida M., Matsuzaki K., Fujimoto R., Mukai K. and 
Osborne J.P., 1997, {\mnras} {287}, 651
\bibitem{met96} Makishima K. {\etal}, 1996, {\pasj} {48}, 171
\bibitem{mtki96} Misaki K., Terashima Y., Kamata Y., Ishida M., Kunieda H. and 
  Tawara Y., 1996, {\apjl} {470}, L53
\bibitem{mc87} McCray R.A., 1987, in Spectroscopy of Astrophysical Plasma, ed.
A.~Dalgarno and D.~Layzer (Cambridge University Press: Cambridge), p.~260
\bibitem{nw89} Norton A. J. and Watson M. G., 1989, {\mnras} {237}, 853
\bibitem{oet96} Ohashi T. {\etal}, 1996, {\pasj} {46}, 157

\bibitem{rmcwc94} Ramsay G., Mason K.O., Cropper M., Watson M.G. and 
Clayton K.L., 1994, {\mnras} {270}, 692

\bibitem{rs77} Raymond J.C., Smith B.W., 1977, {\apjs} {35}, 419

\bibitem{r81} Rothschild R.,  \etal\ 1981, ApJ 250, 723

\bibitem{set95} Serlemitsos P.J. {\etal}, 1995, {\pasj} {47}, 105

\bibitem{ssb97} Shrader C.R., Singh K.P. and Barrett P., 1997, {\apj} {485},
  1006
\bibitem{ss93} Singh J. and Swank J.H., 1993, {\mnras} {262}, 1000
\bibitem{sshf95} Szkody P., Silber A., Hoard D.W., Fierce E., Singh K.P., 
Barrett P., Schlegel E. and Piirola V., 1995, {\apjl} {455}, L43
\bibitem{tr96} Thomas H. -C. and Reinsch K., 1996, {\aaa} {315}, L1
\bibitem{wb96} Woelk U. and Beuermann K., 1996, {\aaa} {306}, 232
\bibitem{wcs94} Wu K., Chanmugam G. and Shaviv G., 1994, {\apj} {426}, 664
\bibitem{yet97} Yamashita A. {\etal}, 1997, {IEEE Trans. Nucl. Sci.} {44}, 847

\end{thebibliography}
\end{document}